# Emergent superconductivity in van der Waals Kagome material $Pd_3P_2S_8$ under high pressure

Qi Wang[1,2,#], Xiaole Qiu[3,#], Cuiying Pei[1], Benchao Gong[3], Lingling Gao[1], Yi Zhao[1], Weizheng Cao[1], Changhua Li[1], Shihao Zhu[1], Mingxin Zhang[1], Yulin Chen[1,2,4], Kai Liu[3,*], Yanpeng Qi[1,2,5,*]

[1]*School of Physical Science and Technology, ShanghaiTech University, Shanghai 201210, China*

[2]*ShanghaiTech Laboratory for Topological Physics, ShanghaiTech University, Shanghai 201210, China*

[3]*Department of Physics and Beijing Key Laboratory of Opto-electronic Functional Materials & Micro-nano Devices, Renmin University of China, Beijing 100872, China*

[4]*Department of Physics, Clarendon Laboratory, University of Oxford, Parks Road, Oxford OX1 3PU, UK*

[5]*Shanghai Key Laboratory of High-resolution Electron Microscopy and ShanghaiTech Laboratory for Topological Physics, ShanghaiTech University, Shanghai 201210, China*

\# These authors contributed to this work equally.
\* Correspondence should be addressed to Y. P. Q. (qiyp@shanghaitech.edu.cn) or K. L. (kliu@ruc.edu.cn)

## Abstract

Kagome lattice systems have been proposed to host rich physics, which provide an excellent platform to explore unusual quantum states. Here, we report on the discovery of superconductivity in van der Waals material $Pd_3P_2S_8$ under pressure. The superconductivity is observed in $Pd_3P_2S_8$ for those pressures where the temperature dependence of the resistivity changes from a semiconducting-like behavior to that of a normal metal. The superconducting transition temperature $T_c$ increases with applied pressure and reaches ~ 6.83 K at 79.5 GPa. Combining high-pressure XRD, Raman spectroscopy and theoretical calculations, our results demonstrate that the observed superconductivity induced by high pressure in $Pd_3P_2S_8$ is closely related to the formation of amorphous phase, which results from the structural instability due to the enhanced coupling between interlayer Pd and S atoms upon compression.

Kagome lattice, which is comprised of corner-sharing triangles and hosts Dirac points, saddle points and flat bands [1-3], has attained considerable attention in the past few years. The materials with kagome lattice display intriguing and unique quantum phenomena related to geometric frustration, nontrivial band topology and strongly correlation, such as quantum spin liquid state, large intrinsic anomalous Hall effect, superconductivity and charge density wave state [4-14]. Since a great many of studies on the kagome superconductor $AV_3Sb_5$ (A = K, Rb, Cs) [11-16], the exploration of superconductivity in novel kagome compounds is promising and intriguing.

$Pd_3P_2S_8$ is a two-dimensional van der Waals (vdW) semiconductor with Pd Kagome lattice [17]. Theoretical study implies that $Pd_3P_2S_8$ possesses the energy gap ~ 1.78 eV and 1.40 eV for monolayer and bulk, respectively. Moreover, it hosts a nearly flat band slightly below Fermi level in the whole Brillouin zone (BZ) when it changes from bulk to monolayer. Recent theoretical work has pointed out that the dispersionless flat bands host large density of states could lead to exotic superconductivity [18]. Therefore, $Pd_3P_2S_8$ offers unique opportunities for the exploration of superconductivity in Kagome lattices.

High pressure is an effective technique without introducing impurities and defects to modulate properties of materials, such as magnetism, superconductivity, structural and topological transition [13, 19-20]. The pressure-induced superconductivity has been frequently investigated in many of topological materials, such as $Bi_2Se_3$, $Bi_4I_4$ etc. [21-23]. In this work, by using high-pressure electrical transport measurements, we found that $Pd_3P_2S_8$ single crystal emerges a superconductivity up to 6.83 K at 79.5 GPa. Combined with the high-pressure x-ray diffraction (XRD), Raman spectroscopy and theoretical calculations, our results suggest that the appearance of superconductivity could be related to the amorphous transition as a result of structural instability arising from the increased coupling of Pd and S atoms between adjacent layers under high pressure.

Single crystals of $Pd_3P_2S_8$ were grown by chemical vapor transport technique [17] with a molar ratio of Pd : P : S = 3 : 2 : 8 in a two-zone furnace. The XRD of $Pd_3P_2S_8$ single crystal at ambient pressure was measured in a Bruker D8 X-ray diffractometer with Cu Kα radiation of wavelength λ = 0.15418 nm. *In situ* high-pressure electrical transport property was performed in physical property measurement system (PPMS-9). The resistivity of $Pd_3P_2S_8$ single crystals was measured by van der Pauw method in a nonmagnetic diamond anvil cell (DAC). In order to make sure electrical insulation, a cubic boron nitride and epoxy mixture layer was employed between BeCu gasket and Pt wires. *In situ* high-pressure XRD measurements were carried out on the beamline BL15U of Shanghai Synchrotron Radiation Facility using x-ray (λ = 0.6199 Å). *In situ* high-pressure Raman spectroscopy experiments were performed using a Renishaw Raman spectrometer (laser excitation wavelength λ = 532 nm). Ruby luminescence method [24] was used for determining the pressure in all high-pressure experiments.

The phonon spectra of $Pd_3P_2S_8$ under pressure were investigated with the density functional perturbation theory [25] calculations as implemented in the Quantum ESPRESSO (QE) package [26]. The kinetic energy cutoffs of the wave functions and the charge densities were chosen to be 60 and 600 Ry, respectively. The Fermi surface was broadened by the Gaussian smearing method with a width of 0.0037 Ry (0.05 eV). In the structural optimization, both lattice constants and internal atomic positions were fully relaxed until the forces on atoms were smaller than 0.0001 Ry/Bohr. A 4 × 4 × 4 **q**-point mesh and a 12 × 12 × 12 **k**-point mesh was used for the Brillouin zone (BZ) sampling of the primitive cell. The DFT-D2 method [27,28] was used to account for the van der Waals interaction between $Pd_3P_2S_8$ layers.

$Pd_3P_2S_8$ is crystalized in hexagonal layered structure with *P*-3*m*1 space group (No. 164) [29]. Pd atoms form perfect two-dimensional kagome lattice [Fig. 1(a)]. In the S-Pd-S slabs marked with dashed rectangular in Fig. 1(a), each Pd atom is surrounded by four S atoms. Moreover, there are partial S atoms exactly lying below and above P atoms. $Pd_3P_2S_8$ single crystal is partially transparent with red. Fig. 1(b) displays the single crystal XRD spectra for $Pd_3P_2S_8$, implying that the surface of $Pd_3P_2S_8$ single crystal is normal to the [001] direction. Our results are in good agreement with the previous report [17] and indicate a high-quality of the single crystals.

We carried out high pressure experiments on $Pd_3P_2S_8$ single crystals to study the effect of pressure on electrical transport. Fig. 2(a) depicts the temperature dependence of resistivity $\rho(T)$ at pressures between 14.3 and 63.2 GPa in run 1. At 14.3 GPa, the resistivity exhibits significant semiconducting behavior, i.e., the $\rho(T)$ curve increases with decreasing temperature. The magnitude of the resistivity initially reaches 0.85 Ω cm at 10 K and 14.3 GPa and declines very fast with further increasing pressures. Surprisingly, at 25.2 GPa, the resistivity begins to drop significantly at low temperature [Fig. 2(b)]. The downward tendency continued until 31.1 GPa. At the pressure of 34.4 GPa, the $\rho(T)$ curve shows metallic behavior, and the resistivity drops to zero at low temperature, indicating the emergence of superconducting state. The $T_{c,onset}$ corresponding to onset superconductivity transition temperature is about 2.82 K, and $T_{c,zero}$ where the resistivity reaches zero is 1.84 K. It is clear that the $T_c$ increases monotonously with increasing pressure. There is no saturated tendency up to 63.2 GPa and the maximum $T_c$ we measured is about 6 K. The transport measurements on different samples for several independent runs provide the consistent and reproducible results [Figs. 2(c)-(d)], confirming this intrinsic superconductivity under pressure. Moreover, the $\rho(T)$ curves as a function of temperature at various fields for 70.1 GPa in run 2 is shown in Fig. 2(e). As the magnetic field increases, the superconductivity is gradually suppressed, accompanied by the reduction in the values of $T_c$. The temperature dependent upper critical field $\mu_0H_{c2}(T)$ is shown in Fig. 2(f). Here, the value of temperature $T$ is derived from the half of normal state resistivity. To determine the upper critical field $\mu_0H_{c2}(0)$ at 0 K, the formula $\mu_0H_{c2}(T) = \mu_0H_{c2}(0)(1-(T/T_c))^{1+\alpha}$ is used to fit the $\mu_0H_{c2}(T)$ curves [13]. The obtained $\mu_0H_{c2}(0)$ for 59.5, 70.1 and 79.5 GPa are 5.6, 6.0 and 6.6 T respectively. These values are all less than the Pauli limiting field 1.84$T_c$.

In order to certify whether the pressure-induced superconductivity is associated with structural phase transition, we performed high pressure XRD studies on $Pd_3P_2S_8$ single crystal. As presented in Fig. 3(a), most Bragg reflections can be nicely refined by using the space group $P$-3$m$1 (No.164) as the initial model. The small peaks marked with asterisks represent the signals of lanthanum gasket, which are attributed to the trailing of X-ray spot. When increasing the pressure, all peaks slowly shift to higher angles. No structural phase transition is observed up to 26.5 GPa. Interestingly, above

30.8 GPa, the strength of the peaks become almost vanished abruptly, except for the peaks of lanthanum. It demonstrates that $Pd_3P_2S_8$ may go through an amorphous phase transition persistent up to 77.1 GPa. In addition, upon decompression, the amorphous behavior is maintained. Moreover, the lattice parameters *a* and *c* determined by means of the structural refinements below 26.5 GPa are presented in Fig. 3(b). They both gradually reduce under pressure and the calculated volume declines with increasing pressure [Fig. 3(c)]. Meanwhile, the Raman spectroscopy experiments were also carried out [Fig. 3(d)]. There is no anomaly below 25.2 GPa, suggesting the absence of structural phase transition. However, above 26.0 GPa, the Raman peaks appreciably weakened, which is very analogous to the phenomenon observed in high-pressure XRD.

To explore the lattice dynamics of $Pd_3P_2S_8$ under pressure, we calculated its phonon dispersions at three typical pressure points 10, 20, and 30 GPa, as shown in Figs. 4 (a)-(c), respectively. Under 10 GPa, the absence of imaginary frequency in the phonon dispersion across the whole BZ indicates the dynamical stability of $Pd_3P_2S_8$. When the pressure increases up to 20 GPa, two of the acoustic branches display imaginary modes around the A point, which means that $Pd_3P_2S_8$ tends to become unstable at this pressure. As the pressure further rises to 30 GPa, almost all symmetry paths in the BZ demonstrate imaginary phonon modes, which corresponds to a large structural distortion of $Pd_3P_2S_8$. The pressure dependence of structural instability derived from our calculations agrees well with our XRD and Raman measurements. If we assume that the lattice framework of $Pd_3P_2S_8$ does not change under pressure, we find that the pressure can induce a gradual reduction of the lattice constants [Fig. 4(d)], especially the one along the *c* direction with the weak interlayer vdW interaction. In contrast to the almost unchanged intralayer Pd-S bond length ($d_{Pd-S}$), the atomic distance between Pd and S atoms from the neighboring layers (*d*) follows exactly the tendency of lattice constant *c* and it even becomes comparable to the intralayer Pd-S bond length at high pressure. The drastic reduction of *d* under high pressure significantly enhances the coupling between Pd and S atoms from the neighboring layers. As a result, the crystal field in which Pd atom resides transits from the original planar tetragonal field to a distorted octahedral field. This may lead to the redistribution of Pd *d* electrons and the John-Teller structural instability.

Fig. 5 illustrates the pressure dependence of both resistivity at 10 K and $T_c$ for $Pd_3P_2S_8$ single crystal in different runs. It can be clearly seen that the resistivity at 10 K decreases rapidly with increasing pressure. The superconductivity ($T_{c,zero}$ ~ 1.84 K) emerges with the pressure applied at 34.4 GPa, where the $\rho(T)$ curve enters into metallic state. Then $T_c$ begins to increase monotonously upon further compression and reaches a maximum value of 6.83 K at 79.5 GPa. As mentioned above, the structure is stable below 26 GPa. However, when the pressure is above about 26 GPa, both high pressure XRD and Raman spectroscopy reveal that the possible occurrence of amorphous transition. The calculated imaginary modes in phonon spectra of $Pd_3P_2S_8$ above 20 GPa are possibly the main reason of the formation of amorphous transition. Interestingly, the superconducting state is also observed in this pressure range. Hence, it indicates that the amorphous transition of $Pd_3P_2S_8$ results in the appearance of superconductivity. Similar superconducting phenomena in the amorphous phases were also observed in other materials, such as $Bi_4I_4$, $(NbSe_4)_2I$, and $MnBi_2Te_4$ [22,30,31], which will stimulate further studies from both experimental and theoretical perspectives.

In summary, the vdW compound $Pd_3P_2S_8$ single crystal with Pd kagome lattice exhibits superconductivity under high pressure, where the temperature dependent resistivity undergoes a semiconducting-metallic behavior transition. With increasing pressure, the superconducting critical temperature $T_c$ is monotonously enhanced to a maximum of 6.83 K at ~ 79.5 GPa. The combined high-pressure XRD, Raman spectroscopy and phonon spectra calculation consistently evidence that the emergence of superconductivity is accompanied with an amorphization, which originates from the instability of structure. The discovery of the superconductivity in $Pd_3P_2S_8$ induced by high pressure provides a platform for the exploration and understanding of superconductivity in novel kagome materials.

Note: When we prepared our manuscript, we learned that two similar works were submitted on the arXiv [32, 33].

**Acknowledgment**

This work was supported by the National Key R&D Program of China (Grant No. 2018YFA0704300, 2017YFA0302903), the National Natural Science Foundation of China (Grant No. U1932217, 11974246, 12174443), the Natural Science Foundation of Shanghai (Grant No. 19ZR1477300), the Science and Technology Commission of Shanghai Municipality (19JC1413900), Shanghai Science and Technology Plan (Grant No. 21DZ2260400), the Beijing Natural Science Foundation (Grant No. Z200005), and the China Postdoctoral Science Foundation (Grant No. 2021M692132). The authors thank the support from Analytical Instrumentation Center (# SPST-AIC10112914), SPST, ShanghaiTech University. The authors thank the staffs from BL15U1 at Shanghai Synchrotron Radiation Facility for assistance during data collection. Computational resources have been provided by the Physical Laboratory of High-Performance Computing at Renmin University of China.

**Reference**

[1] M. Kang, L. Ye, S. Fang, J.-S. You, A. Levitan, M. Han, J. I. Facio, C. Jozwiak, A. Bostwick, E. Rotenberg, M. K. Chan, R. D. McDonald, D. Graf, K. Kaznatcheev, E. Vescovo, D. C. Bell, E. Kaxiras, J. v. d. Brink, M. Richter, M. P. Ghimire, J. G. Checkelsky, and R. Comin, Nat. Mater. 19, 163 (2020).

[2] M. Kang, S. Fang, L. Ye, H. C. Po, J. Denlinger, C. Jozwiak, A. Bostwick, E. Rotenberg, E. Kaxiras, J. G. Checkelsky, and R. Comin, Nat. Commun. 11, 4004 (2020).

[3] M. Kang, S. Fang, J.-K. Kim, B. R. Ortiz, S. H. Ryu, J. Kim, J. Yoo, G. Sangiovanni, D. D. Sante, B.-G. Park, C. Jozwiak, A. Bostwick, E. Rotenberg, E. Kaxiras, S. D. Wilson, J. -H. Park, and R. Comin. Nat. Phys. 18, 301 (2022).

[4] L. Balents, Nature 464, 199 (2010).

[5] Y. Zhou, K. Kanoda, and T.-K. Ng, Rev. Mod. Phys. 89, 025003 (2017).

[6] S. Nakatsuji, N. Kiyohara, and T. Higo, Nature 527, 212 (2015).

[7] L. Ye, M. Kang, J. Liu, F. von Cube, C. R. Wicker, T. Suzuki, C. Jozwiak, A. Bostwick, E. Rotenberg, D. C. Bell, L. Fu, R. Comin, and J. G. Checkelsky, Nature 555, 638 (2018).

[8] Q. Wang, Y. Xu, R. Lou, Z. Liu, M. Li, Y. Huang, D. Shen, H. Weng, S. C. Wang, and H. C. Lei, Nat. Commun. 9, 3681 (2018).

[9] J.-X. Yin, W. Ma, T. A. Cochran, X. Xu, S. S. Zhang, H.-J. Tien, N. Shumiya, G. Cheng, K. Jiang, B. Lian, Z. Song,G. Chang, I. Belopolski, D. Multer, M. Litskevich, Z.-J. Cheng, X. P. Yang, B. Swidler, H. Zhou, H. Lin, T. Neupert, Z. Wang, N. Yao, T.-R. Chang, S. Jia, and M. Z. Hasan, Nature 583, 533 (2020).

[10] L. Gao, S. Shen, Q. Wang, W. Shi, Y. Zhao, C. Li, W. Cao, C. Pei, J.-Y. Ge, G. Li, J. Li, Y. Chen, S. Yan, and Y. Qi, Appl. Phys. Lett. 119, 092405 (2021) .

[11] B. R. Ortiz, S. M. L. Teicher, Y. Hu, J. L. Zuo, P. M. Sarte, E. C. Schueller, A. M. M. Abeykoon, M. J. Krogstad, S. Rosenkranz, R. Osborn, R. Seshadri, L. Balents, J. He, and S. D. Wilson, Phys. Rev. Lett. 125, 247002 (2020).

[12] Q. W. Yin, Z. J. Tu, C. S. Gong, Y. Fu, S. H. Yan, and H. C. Lei, Chin. Phys. Lett. 38, 037403 (2021).

[13] Q. Wang, P. Kong, W. Shi, C. Pei, C. Wen, L. Gao, Y. Zhao, Q. Yin, Y. Wu, G. Li,

H. C. Lei, J. Li, Y. L. Chen, S. C. Yan, and Y. P. Qi, Adv. Mater. 2102813 (2021).

[14] Y.-X. Jiang, J.-X. Yin, M. M. Denner, N. Shumiya, B. R. Ortiz, G. Xu, Z. Guguchia, J. He, M. S. Hossain, X. Liu, J. Ruff, L. Kautzsch, S. S. Zhang, G. Chang, I. Belopolski, Q. Zhang, T. A. Cochran, D. Multer, M. Litskevich, Z.-J. Cheng, X. P. Yang, Z. Q. Wang, R. Thomale, T. Neupert, S. D. Wilson, and M. Z. Hasan, Nat. Mater. 10, 1038 (2021).

[15] Y. Fu, N. Zhao, Z. Chen, Q. Yin, Z. Tu, C. Gong, C. Xi, X. Zhu, Y. Sun, K. Liu, and H. C. Lei, Phys. Rev. Lett. 127, 207002 (2021).

[16] Y. Xiang, Q. Li, Y. Li, W. Xie, H. Yang, Z. Wang, Y. Yao, and H.-H. Wen, Nat. Commun. 12, 6727 (2021).

[17] S. Park, S. Kang, H. Kim, K. H. Lee, P. Kim, S. Sim, N. Lee, B. Karuppannan, J. Kim, J. Kim, K. I. Sim, M. J. Coak, Y. Noda, C.-H. Park, J. H. Kim, and J.-G. Park, Sci. Rep. 10, 20998 (2020).

[18] W.-H. Ko, P. A. Lee, and X.-G. Wen, Phys. Rev. B 79, 214502 (2009).

[19] B. Wang, Y. Liu, K. Ishigaki, K. Matsubayashi, J. Cheng, W. Lu, Y. Sun, and Y. Uwatoko, Phys. Rev. B 95, 220501(R) (2017).

[20] T. Ideue, M. Hirayama, H. Taiko, T. Takahashi, M. Murase, T. Miyake, S. Murakami, T. Sasagawa, and Y. Iwasa, PNAS 116, 25530 (2019).

[21] K. Kirshenbaum, P. S. Syers, A. P. Hope, N. P. Butch, J. R. Jeffries, S. T. Weir, J. J. Hamlin, M. B. Maple, Y. K. Vohra, and J .Paglione, Phys. Rev. Lett. 111, 087001 (2013).

[22] Y. Qi, W. Shi, P. Werner, P. G. Naumov, W. Schnelle, L. Wang, K. G. Rana, S. Parkin, S. A. Medvedev, B. Yan, and C. Felser, npj Quant Mater 3, 4 (2018).

[23] F. H. Yu, X. Y. Hua, T. Chen, J. Sun, M. Z. Shi, W. Z. Zhuo, D. H. Ma, H. H. Wang, J. J. Ying, and X. H. Chen, New J. Phys. 22, 123013 (2020).

[24] H. K. Mao, J. Xu, and P. M. Bell, J. Geophys. Res. 91, 4673 (1986).

[25] S. Baroni, S. de Gironcoli, A. Dal Corso, and P. Giannozzi, Rev. Mod. Phys. 73, 515 (2001).

[26] P. Giannozzim, S. Baroni, N. Bonini, M. Calandra, R. Car, C. Cavazzoni, D. Ceresoli, G. L. Chiarotti, M. Cococcioni, and I. Dabo *et al.*, J. Phys.: Condens. Matter 21, 395502 (2009).

[27] S. Grimme, J. Comp. Chem. 27, 1787 (2006).

[28] V. Barone, M. Casarin, D. Forrer, M. Pavone, M. Sambi, A. Vittadini, J. Comp. Chem. 30, 934 (2009).

[29] T. A. Bither, P. C. Donohue, and H. S. Young, J. Solid State Chem. 3, 300 (1971).

[30] C. Pei, W. Shi, Y. Zhao, L. Gao, J. Gao, Y. Li, H. Zhu, Q. Zhang, N. Yu, C. Li, W. Cao, S. A. Medvedev, C. Felser, B. Yan, Z. Liu, Y. Chen, Z. Wang, and Y. Qi, Mater. Today Phys. 21, 100509 (2021).

[31] C. Pei, Y. Xia, J. Wu, Y. Zhao, L. Gao, T. Ying, B. Gao, N. Li, W. Yang, D. Zhang, H. Gou, Y. Chen, H. Hosono, G. Li, and Y. Qi, Chin. Phys. Lett. 37, 066401 (2020).

[32] S. Li, S. Han, S. Yan, Y. Cui, L. Wang, S. Wang, S. Chen, H. Lei, F. Yuan, J. Zhang, and W. Yu, arXiv:2203.15957.

[33] Y. Zhou, X. He, S. Wang, J. Wang, X. Chen, Y. Zhou, C. An, M. Zhang, Z. Zhang, and Z. Yang, arXiv:2203.16943.

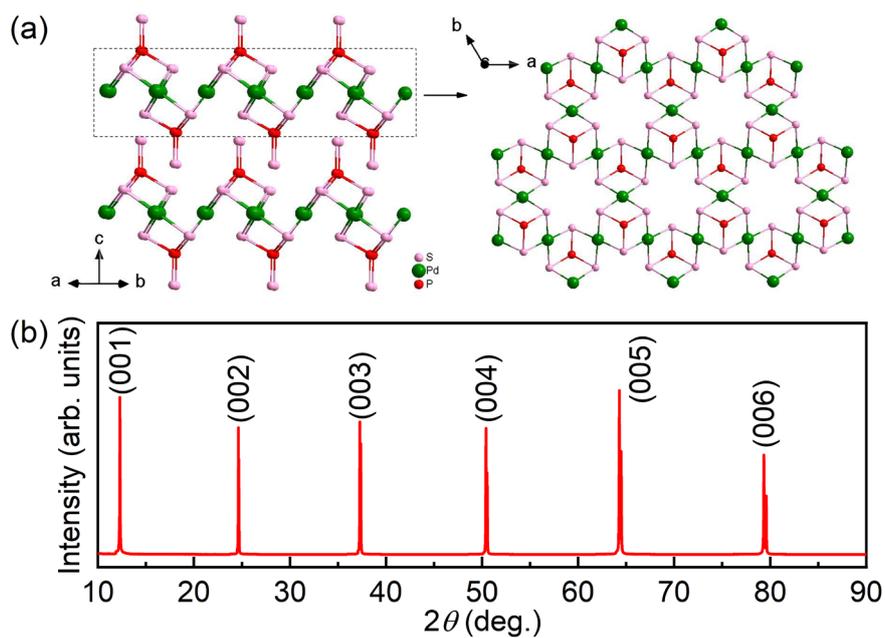

Fig. 1. (a) A schematic of crystal structure of $Pd_3P_2S_8$ (left) and the top view of Pd kagome layer (right). Green, red and pink spheres denote Pd, P and S atoms, respectively. (b) XRD pattern of a $Pd_3P_2S_8$ single crystal.

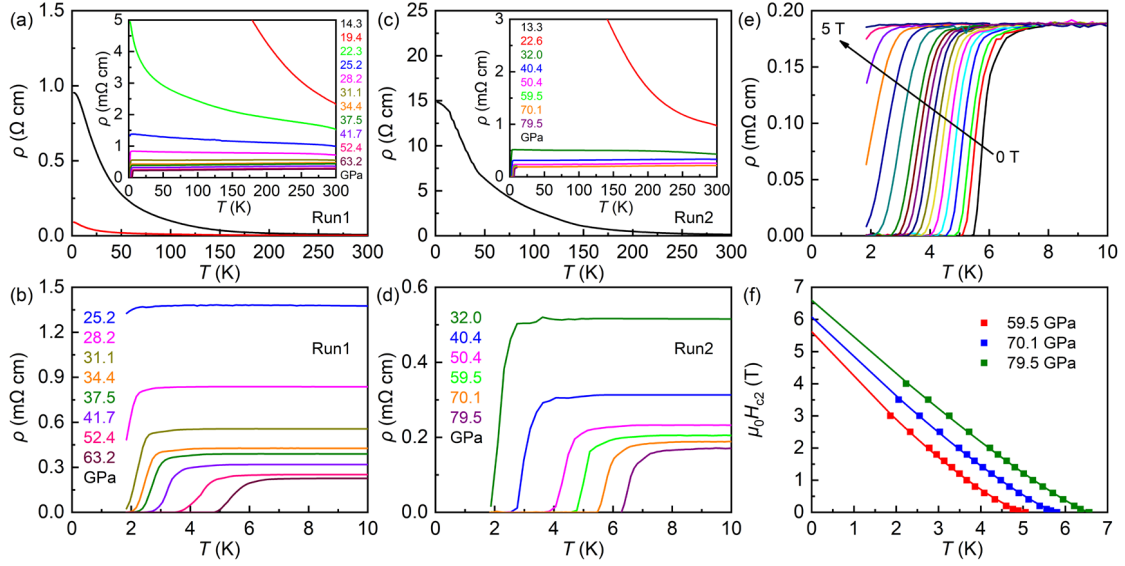

Fig. 2. (a) The resistivity $\rho(T)$ as a function of temperature under various pressures for run 1. (b) Temperature dependence of resistivity $\rho(T)$ below 10 K from 25.2 to 63.2 GPa. (c) Temperature dependence of $\rho(T)$ at different pressures for run 2. (d) Temperature dependent $\rho(T)$ below 10 K from 32.0 to 79.5 GPa. (e) Temperature dependence of resistivity $\rho(T)$ at different fields for 70.12 GPa. (f) The upper critical field $\mu_0 H_{c2}(T)$ as a function of temperature at representative pressures. The solid lines correspond to the results of fitting by means of the formula $\mu_0 H_{c2}(T) = \mu_0 H_{c2}(0)(1-(T/T_c))^{1+\alpha}$.

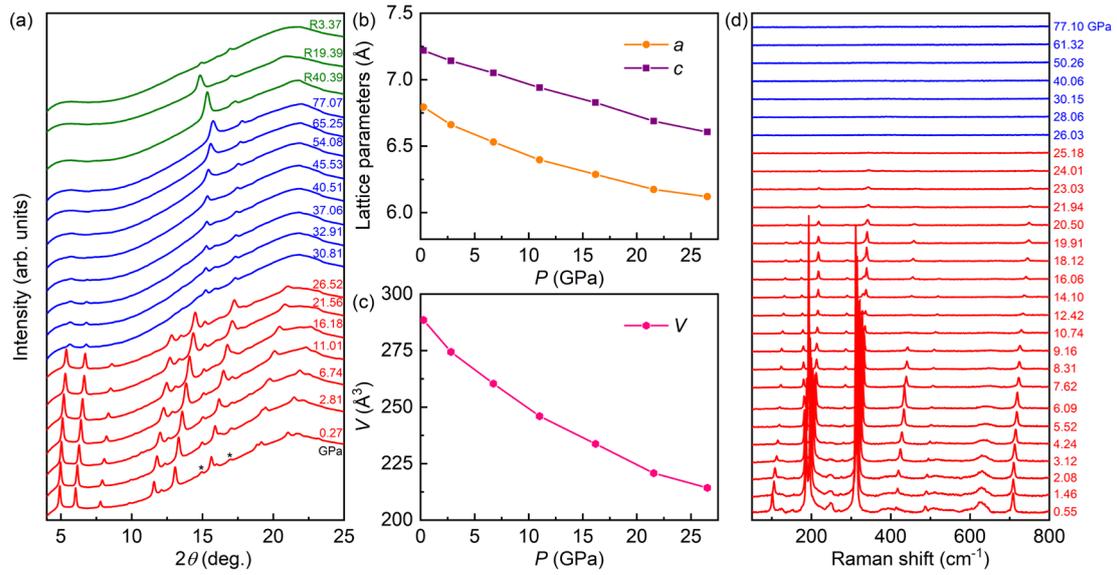

Fig. 3. (a) Powder XRD pattern for Pd$_3$P$_2$S$_8$ single crystal under various pressures at room temperature. The asterisks denote the signals of lanthanum gasket. (b) and (c) show the pressure-dependent lattice parameters $a$, $c$ and volume determined from refinements. (d) Raman spectra of Pd$_3$P$_2$S$_8$ at various pressures measured at room temperature.

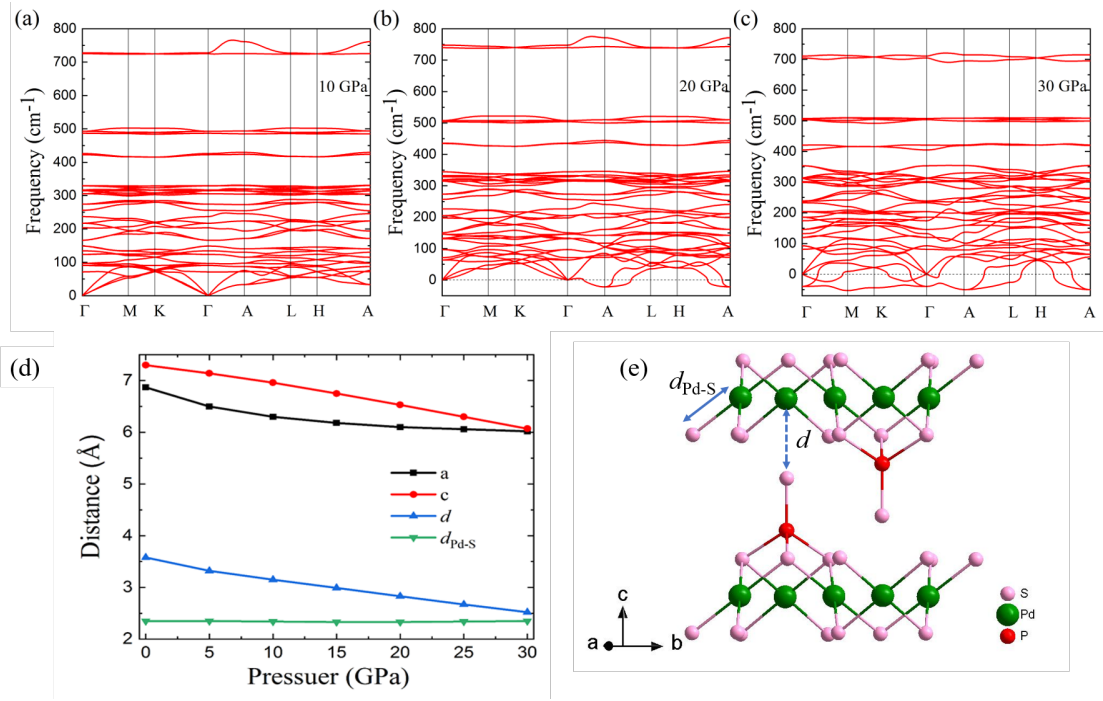

Fig. 4. Phonon spectra of $Pd_3P_2S_8$ at (a) 10 GPa, (b) 20 GPa and (c) 30 GPa, respectively. (d) The evolutions of lattice constants *a* and *c*, intralayer bond length between Pd and S atoms $d_{Pd-S}$, and interlayer distance between Pd and S atoms *d* under pressure. The grey, violet and yellow balls denote the Pd, P and S atoms, respectively. The definitions of atomic distances $d_{Pd-S}$ and *d* are displayed in panel (e).

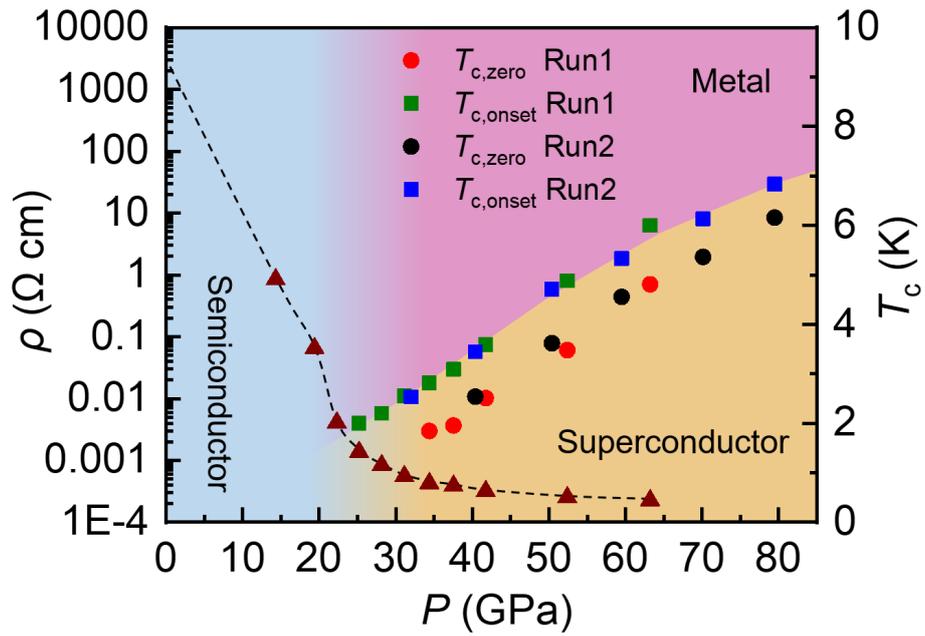

Fig. 5. The phase diagram of Pd$_3$P$_2$P$_8$ as determined from the $\rho(T)$ curves for different runs. Wine triangles represent the values of resistivity at 10 K. Red and black circles represent $T_{c,onset}$ for run1 and run2. On the other hand, green and blue squares represent $T_{c,zero}$ for run1 and run2.